# Application of Model Derived Charge Transfer Inefficiency Corrections to STIS Spectroscopic CCD Data

P. Bristow

*Abstract*

*The ST-ECF Calibration Enhancement effort for the Space Telescope Imaging Spectrograph (STIS) aims to improve data calibration via the application of physical modelling techniques. As part of this effort we have developed a model of the STIS CCD readout process. The model itself is described in some detail in earlier ISRs. Here we consider the application to STIS spectroscopic data. We find good agreement between the simulation derived corrections and empirical corrections for a range of background and source levels. Moreover we find that, where the two differ, because only the simulation properly accounts for the charge distribution in the array.*

# 1. Introduction

Charge Coupled Devices (CCDs) operating in hostile radiation environments suffer a gradual decline in their Charge Transfer Efficiency (CTE, Charge Transfer Inefficiency, CTI=1-CTE). The Space Telescope Imaging Spectrograph (STIS) and the Wide Field and Planetary Camera 2 (WFPC2) on HST have both had their CTE monitored during their operation in orbit and both indeed show a measurable decline in CTE which has reached a level which can significantly affect scientific results (e.g. Cawley et al 2001, Heyer 2001, Kimble, Goudfrooij and Gililand 2000).

Instead of seeking empirical corrections we have chosen to construct a physical model which allows us to predict the effects of degraded CTE in astrophysical data using STIS as a test case. Detailed discussion of the model development and the physics involved can be found in Bristow & Alexov (2002) and Bristow 2003a. Here we restrict the discussion to a review of the performance of the model when used to correct STIS spectroscopic data, particularly in comparison to the empirical correction algorithm. (Bristow 2003b presents a similar discussion regarding STIS imaging data).

Whilst there exists already in the literature a great deal of discussion of the effects of CTI upon astronomical imaging data, very little is said about effects in spectroscopic data. In the case of HST this is simply because STIS is the first spectroscopic instrument to use a CCD and hence the first to suffer from CTI. Moreover, in most cases, the CTI effects in spectroscopic data can often be mitigated by careful planning of the observation so that the spectra from the object of interest falls close to the readout register (Kimble, Goudfrooij and Gililand 2000). Nevertheless, there exist data for which this was either not possible or simply not done and, as we shall see below, there are cases where even well planned data may still suffer from CTI degradation.

Bohlin and Goudfooij (hereafter BG2003) present a detailed analysis of CTI effects in STIS spectroscopic data specifically obtained for this purpose and derive an empirical algorithm for correcting CTE loss in spectrometry of point sources. As discussed in Bristow & Alexov (2002) and Bristow (2003a), empirical corrections for STIS provide both a source of fine calibration and a benchmark for the performance of our rather more expensive (in terms of CPU time and user interaction) model based corrections.

Empirical corrections give the fractional CTE loss as a function of signal strength (for a point source), background, epoch and position on the chip (distance from the readout register). Spectra are assumed to be dispersed along the x axis, perpendicular to the readout direction. It is necessary to formulate and calibrate the corrections differently for photometric and spectroscopic data because of the differing nature of the flux (and therefore charge) distributions. Moreover, *existing empirical corrections only apply to point sources and for a specific extraction procedure* (in this case a 7 pixel high aperture). By modelling the readout process we are able to correct for any charge distribution and therefore any spectral extraction procedure can be applied and extended sources are corrected just as well as point sources.

However, we should expect that the model based corrections are in general agreement with the empirical corrections for point sources. Indeed, by demanding such a general agreement we can use the empirical corrections to calibrate the physical model. This is much easier than returning to, and re-analysing, the data itself as the empirical corrections are essentially a distillation of what is to be learned from the data with respect to CTI. That is not to say that the physical model is simply constrained to reproduce the empirical corrections, this is only the case for point sources and only on average. Instances of disagreement are interesting as they will tell us if the model based correction is either failing (if we can see no good reason for the disagreement) or improving upon the empirical corrections by predicting differences in the charge distribution which empirical corrections could not have dealt with.

## 2. Data, Reduction and Analysis

As part of their effort to monitor STIS CTI, the Spectrographs group at STScI has made observations specifically designed to allow measurement of CTI in spectroscopic data (proposal 8911 Guodfrooij). These observations consist of exposures of identical fields read out through the B and D amplifiers respectively. These correspond to opposite registers meaning that a signal at the top of the chip which is read out through the top register (and experiences very little CTI degradation) can be compared to its CTI affected equivalent read out through the bottom register. The inverse is clearly true for signals at the bottom of the chip.

The datasets which we consider in the analysis here are listed in table 1 along with their exposure times. All are observations of the same field in NGC 346 (Target Ra, Dec: 00 59 00.70, -72 09 59.2), using grating G430L and obtained on 30th August 2001.

| Dataset name | No. Exposures | Duration (s) | Amplifier |
| --- | --- | --- | --- |
| o6if02010 | 10 | 20.0 | D |
| o6if02020 | 4 | 100.0 | D |
| o6if02030 | 2 | 491.5 | D |
| o6if02040 | 10 | 20.0 | D |
| o6if02050 | 4 | 100.0 | D |
| o6if020g0 | 2 | 491.5 | B |
| o6if020i0 | 4 | 100.0 | B |
| o6if020j0 | 2 | 491.5 | B |
| o6if020k0 | 9 | 20.0 | B |

In order to simulate the readout of these datasets we want to work with the best possible estimate of what the charge distribution on the chip was before readout. This problem is discussed in some detail in Bristow (2003a). The best that we can do is set the BLEVCORR calibration switch to PERFORM and all others to OMIT and run CALSTIS. This results in the appropriate bias level being subtracted and the array being cropped to remove overscan areas. For details of the readout simulation see Bristow 2003a. We have used here the optimum parameters presented therein. As part of the pipeline scripts for running our model (see Bristow 2003a) we include an automated calculation of the BG2003 corrections, this is used to derive the points in our plots in section 4.

The process is then:

1. The exposure is scanned and the rows containing the highest signal in a seven pixel tall aperture (as BG2003), presumably containing the spectrum of interest, are found. Cosmic rays and hot pixels which, for short exposures, could lead to false identification of the rows of interest, are excluded by summing the minimum value pixel in each horizontal group of four.

2. The backgrounds to be used for each column are derived by smoothing the pixel array and finding the minimum pixel value between the spectrum and the readout amp. We note that choosing the minimum value is not what one might expect from CTE trapping theory (e.g. Kimble et al 2000), however we follow BG2003 here as they report that this gave the best results for their empirical correction.

3. For each bin of the spectrum (which contains the counts from the seven pixel high extraction), we compute the empirical correction using the distance from the readout register, signal and background as computed above and the date of the dataset.

4. We also extract the same seven rows from the simulation corrected data. This time, for each bin we compute the ratio of uncorrected to corrected counts.

5. In some cases we also scanned the equivalent exposure readout from the opposite amplifier and computed the ratio of the signal in each. We did this only for exposures with bright spectra near to the B amplifier and computed the ratio D counts to B counts.

## 3. Comparison of individual datasets

Figure 1. shows the brightest stellar spectrum in the raw data (i.e. a single exposure) of datasets O6IF02020 (B amp 100s) and O6IF020I0 (D amp 100s) which is located close to the bottom of the array (i.e. close to the B amp). 1a is the spectrum as it appears in the B amp data. The charge packets from these pixels have had to undergo very few charge transfers. By contrast the same spectrum in the D amp data (figure 1b) is represented by charge packets which have been transferred out across the full height of the array. Clearly visible are trails

underneath hot pixels and cosmic rays, in particular hot pixels which are common between the two images have trails in 1b, but not in 1a. Also apparent is the extended glow under the spectrum in 1b.

Figure 1a: Section of the raw image array of dataset O6IF02020 which experiences very little CTI on this part of the chip. Dispersion is along the horizontal axis.

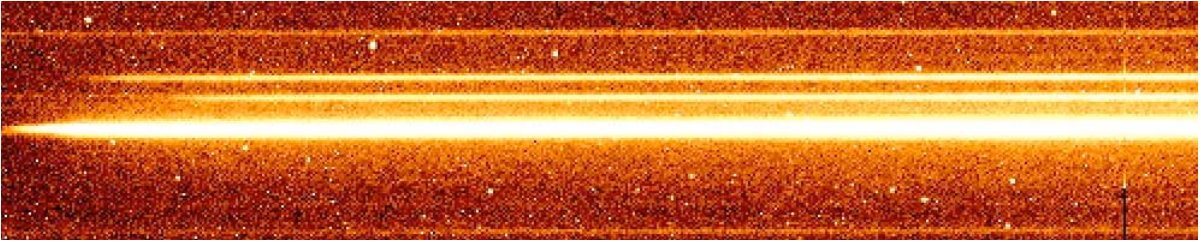

Figure 1b: As 1a but for O6IF020I0 which experiences significant CTI on this part of the chip.

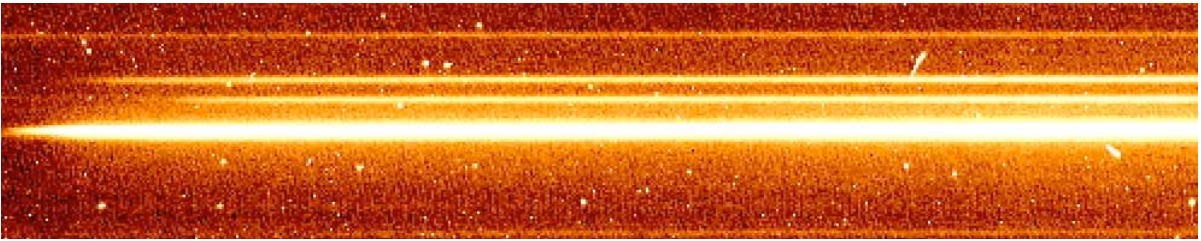

Figure 1c: The raw data of 1b after correction with our model.

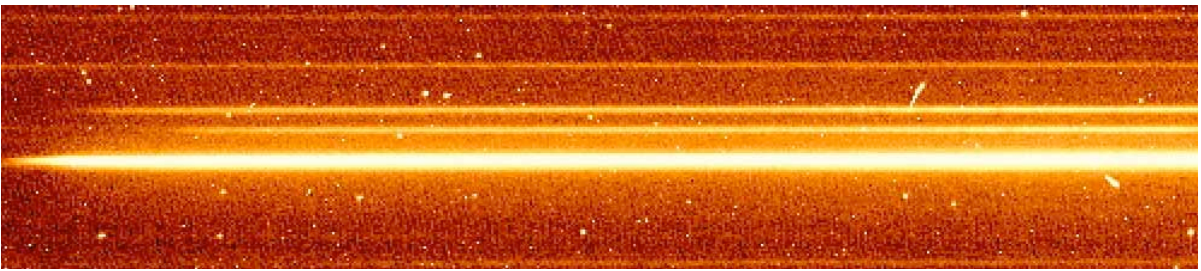

We might hope to see directly the CTI effect in this spectrum by subtracting O6IF02020 from O6IF020I0. However, in practise, this is complicated by slight differences in the readout electronics of the two amplifiers and more importantly, the uncertainty in the exact offset of the two images. Only a very small relative offset in the y direction (perpendicular to the dispersion) is required to obliterate the CTI effects.

Figure 1c shows the spectrum from O6IF02020 after application of the model based correction. The trails and glow under the spectrum are much reduced, though not completely removed. A direct comparison to the spectrum of 1a is not appropriate, once again due to the uncertainties in offset and readout electronics.

In figure 2 we plot the average (over 200 columns) vertical cross section of the spectrum for the B amplifier data (solid line), D amplifier data (dashed line) and the simulation corrected D amplifier data (dotted line and offset for clarity). The attenuation of the signal is clear in the D amplifier data as is its restoration in the corrected data.

Figure 2: Vertical cross section of the image array shown in figure 1 averaged over 200 columns. Dashed line corresponds to 1a; solid line corresponds to 1b; dotted line (offset) corresponds to 1c.

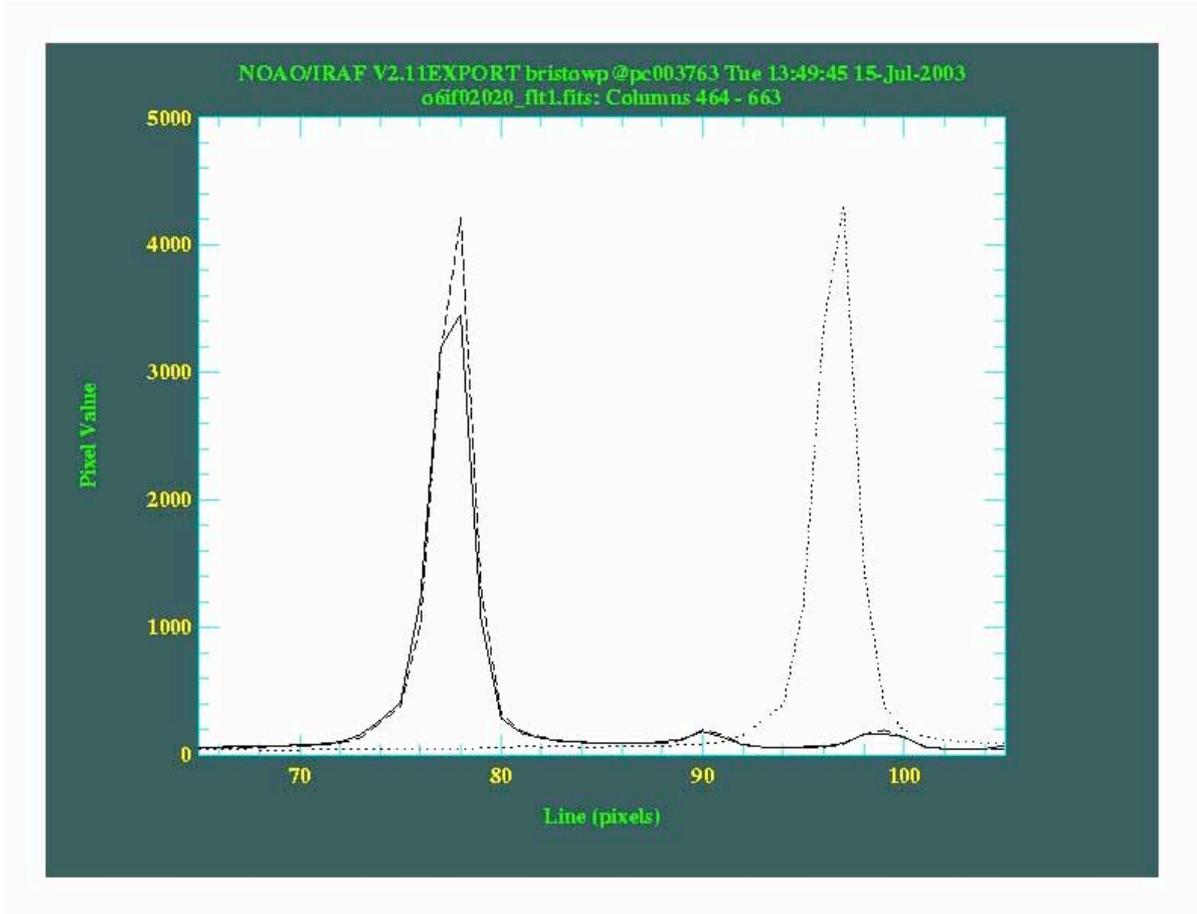

Figure 3 shows the ratio of charge measured in the D amp data to that of B amp data plotted against the fraction of charge left in the D amp data for a spectrum read out from near the bottom of the chip according to the BG2003 corrections, for the same spectrum. This was derived as described in section 2 above, the ordinate values being the results of step 3 and the abscissa values the results of step 5 (using O6IF02030 D amp data and O6IF020G0 B amp data). Immediately apparent is that although the points do lie around x=y there is enormous scatter. This is due to the shot noise difference between the two datasets. It is for this reason that in what follows we make our comparisons to the empirical corrections which effectively contain the information from many datasets.

Figure 3: Comparison between empirical correction and the ratio of signals in the bins of the spectrum extracted from a B and D amplifier exposure.

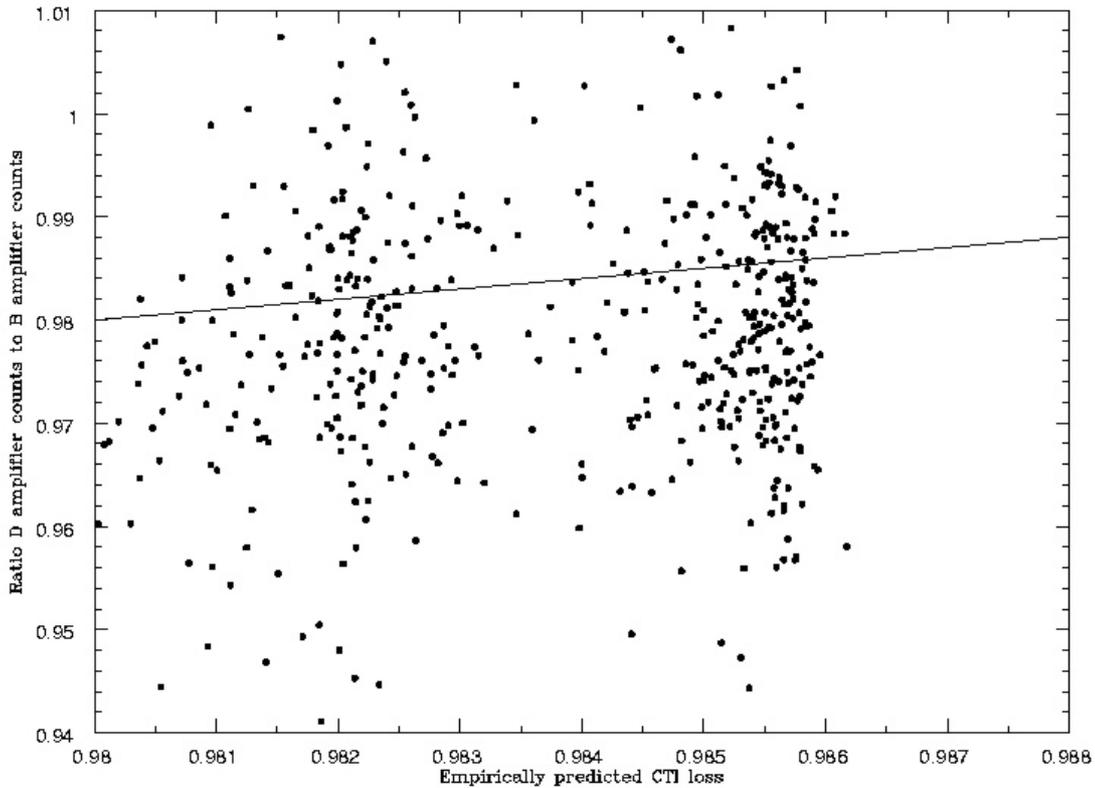

Figures 1, 2 and 3 serve to illustrate the nature of CTI effects in raw data. However, we need to overcome two problems in order to perform a more thorough evaluation of the simulation derived corrections. Firstly, we clearly need to analyse more than just one dataset and ideally sample different background and signal levels. Secondly, as noted above, very slight offsets in the vertical alignment of any pair of B and D amp exposures distort the CTI signal. In order to circumvent this problem we would need to drizzle the exposures together onto a higher resolution grid with the correct jitter offsets. Both of these problems are compounded by the fact that the uncorrelated shot noise between any pair of B and D amp exposures masks the signal of CTI effects in the real data whilst the noise in the simulation corrected data is correlated with that in the corresponding uncorrected exposure.

Therefore, in what follows we will make use of the empirical corrections of Kimble & Goudfrooij (2003). This provides us with a fit to the average level of CTI found in a large number of datasets covering a range of exposure times. We can then deal with relatively few datasets. However, the application of these corrections leaves some room for interpretation.

# 4. Comparison Between Empirical and Simulation Derived Corrections

Figure 4 is similar to figure 3 except that now the ordinate values represent the ratio of counts in the raw data to those in the simulation corrected data for each bin of the extracted spectra. The data in figure 4 come from a number of exposures covering a range of background and signal levels. Perfect agreement between the two estimates of CTI for each bin of the spectrum would result in all points lying on $x=y$. Note that lower exposure times correspond to higher CTI (and therefore lower values on both axis of this plot) because the background is lower. In figure 4 we see three groups of points corresponding to 3 exposure times. Within the groups there are points from bins which have a range of signal levels. Higher signal levels imply (lower CTI and therefore) higher values in this plot.

Figure 4: Comparison between simulation derived corrections and empirical corrections (see text). The box at the top right represents the area plotted in figure 5a below

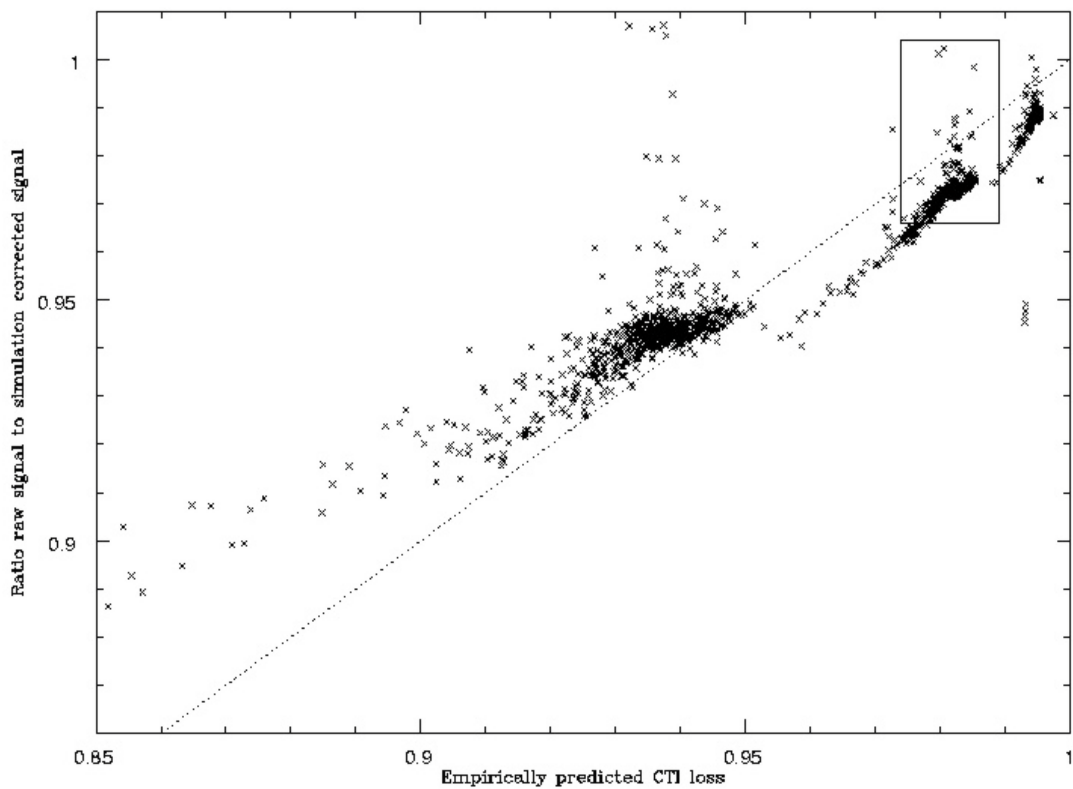

The general agreement of the simulation derived corrections with the empirical corrections over a range of background levels and signal strengths is certainly reassuring. The empirical corrections represent the best fit to real data for point source spectra so we require that on average the simulation estimates are consistent. However, if the agreement was perfect, then whilst this may seem to be confirmation of the accuracy of the model, it would also suggest that the process was one which could be accurately described by an empirical fit even down to the individual pixel level. In this case there would be no need for the application of a more sophisticated model.

In fact we see quite a scatter in the simulation values. This is not the same as the scatter in figure 3 which was mainly due to the differing shot noise between the two exposures. Here, the comparison is between the corrected and uncorrected pixel arrays for which the shot noise is correlated (though there is a relatively small amount of noise added by the simulation of the readout process, see Bristow 2003a). In this case the scatter is to be expected as the simulation has taken into account the charge distribution in, and trapping history of, all of the pixels read out ahead of those in the extraction area.

For example, if we consider the outliers highlighted in the close up section of figure 4 shown in figure 5a and we identify the parts of the spectrum on the raw data to which they correspond, then we see that there are good reasons for their discrepant values. For example, the point labelled $\alpha$ in figure 5a corresponds to the spectrum bin highlighted in figure 5b. Clearly visible directly above this bin is a bright pixel which turns out to be a hot pixel. Figure 5c is an image of the difference between the simulation corrected data and the raw data. Pixels which have gained a lot of charge in the correction (i.e. which would have lost a lot in the read out process) are dark, those which have lost a lot are light. As expected the hot pixel appears dark, but more importantly, beneath is a trail of pixels which would have gained charge, which continues into the extraction area. During the readout process, charge collected in the spectrum pixels would always be being transferred up the chip behind the charge trail of the hot pixel. The state of the charge traps in the pixels traversed by the spectrum charge when it arrived would be quite different from the usual case of low background in the column immediately above the spectrum bin. This leads to reduced attenuation and even accretion of charge left by the hot pixel. For this reason point $\alpha$ lies to the right of the main group in figure 5a. Points $\beta$ and $\gamma$ are actually adjacent bins in the spectrum and have a similar explanation to that of point $\alpha$, the only difference being that in this instance the CTI trail came from a cosmic ray which affected the two columns, not a hot pixel. This can be seen in figures 5d and 5e, whilst figures 5f and 5g illustrate the identical cause of points $\delta$ and $\varepsilon$.

Figure 5 a) Detail from figure 4 with some of the outlying points labelled. b) 7 pixel tall extraction corresponding to point α, in 5a as it appears in the raw data. c) 7 pixel tall extraction corresponding to point α, in 5a as it appears in the difference image generated by the simulation. d) As 5b but for points β (left) and γ (right). e) As 5c but for points β (left) and γ (right). f) As 5b but for points δ (right) and ε (left). g) As 5c but for points δ (right) and ε (left).

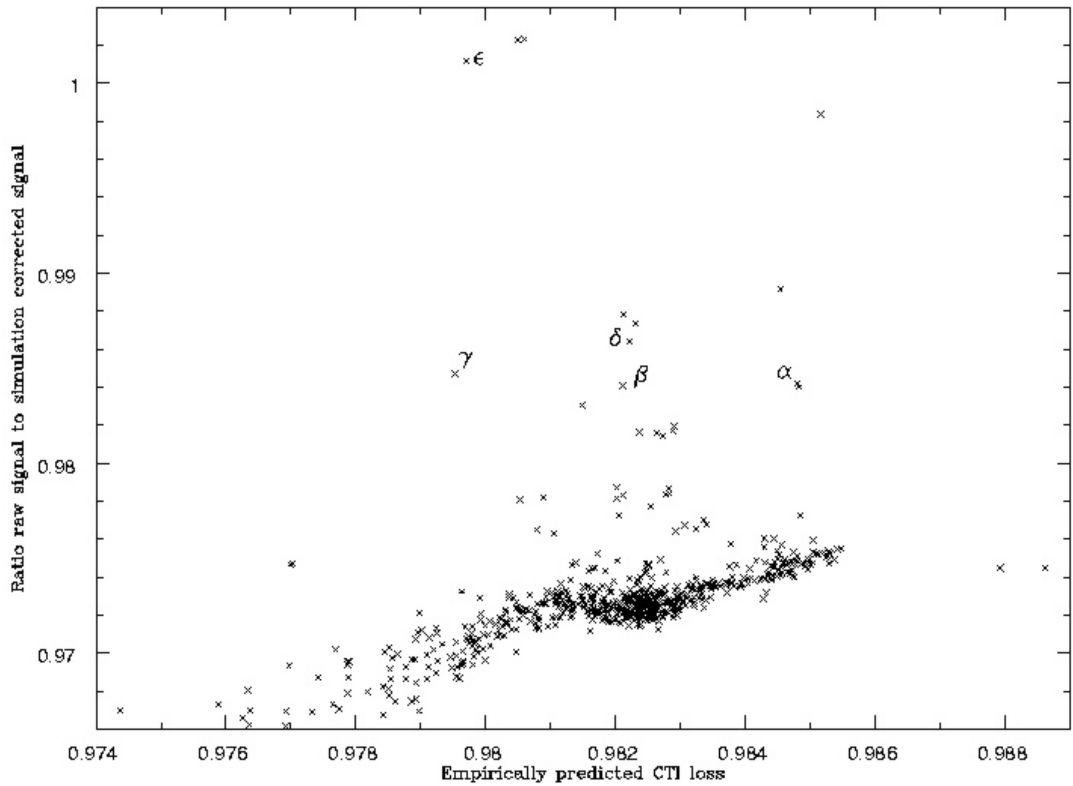

a

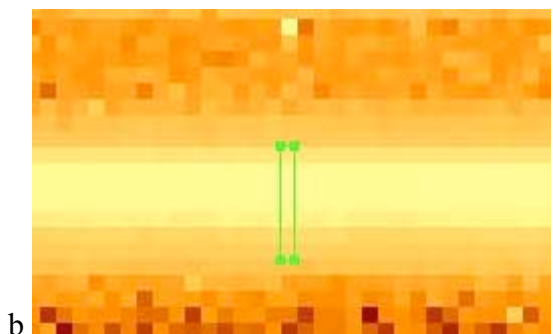

b

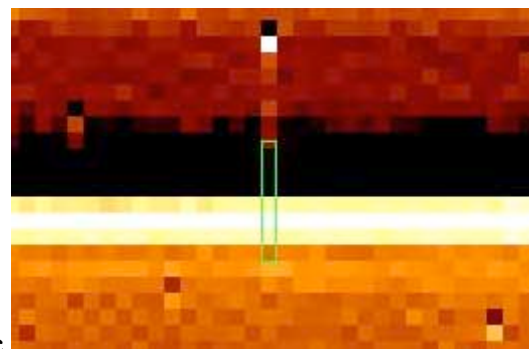

c

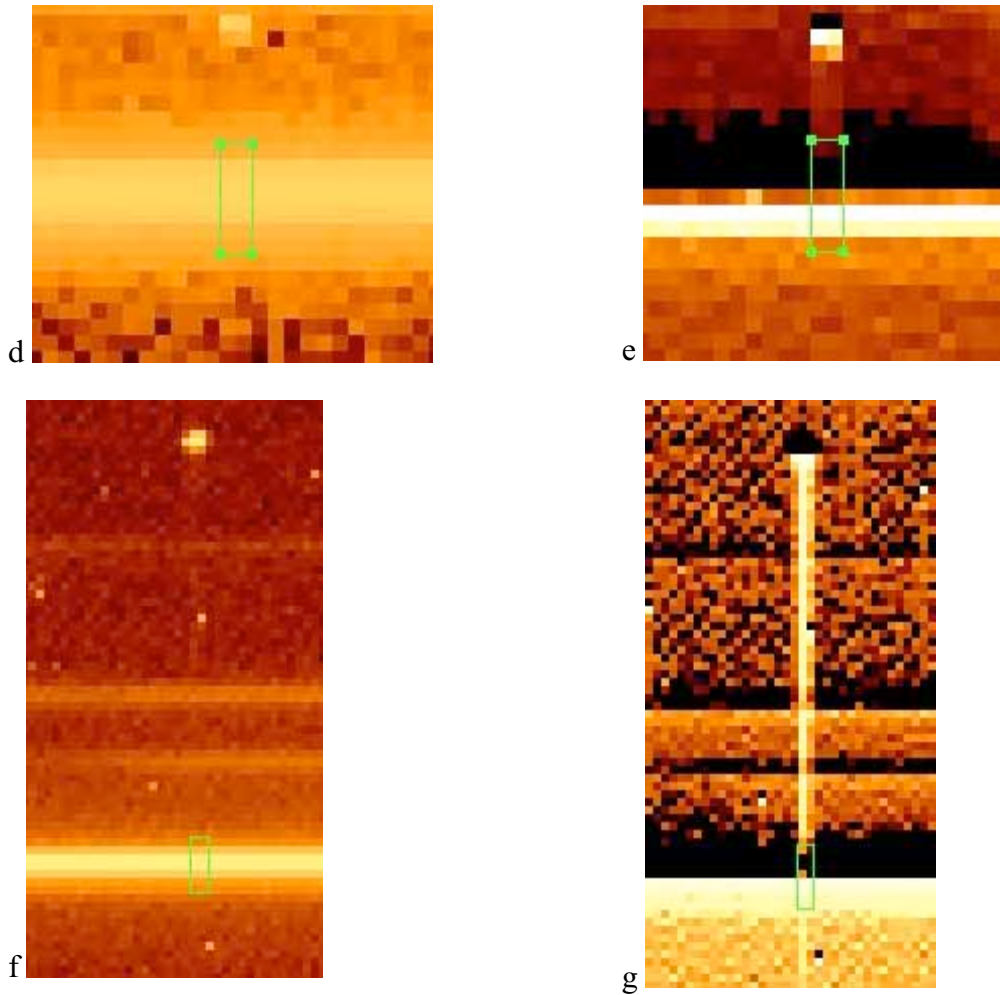

Moreover, in the pipeline calibrated 2D data the cosmic rays and hot pixels which have been smeared into the spectrum of interest will have been removed. Their CTE trails will however remain, but at a level which is not obvious and not likely to alert the conscientious observer who took the time to check the 2D data, of the potential of empirical corrections being inappropriate.

Analysis of other outliers sometimes fails to turn up such an obvious reason for the anomalous values. However, the readout process is rather complex and we should not expect that we can always see an obvious cause and effect. This is exactly why we need a complex simulation approach to properly correct for readout effects.

The scatter in figure 4 can therefore be attributed to more subtle effects of the charge distribution in the readout process. There is, nevertheless some systematic deviation from $x=y$ in this plot however. The simulation derived corrections tend to underestimate the empirical ones at shorter exposure times and overestimate at longer exposure times. The effect suggests a weaker dependence upon background level in the simulations. However, as already noted, there is some room for interpretation in the application of the empirical corrections. The choice of the lowest background level in the pixels read out before the spectrum is rather arbitrary. These uncertainties render the relatively small discrepancy in figure 4 insignificant.

The simulation derived corrections can be refined by running subsequent iterations of the simulation (see Bristow & Alexov 2002 and Bristow 2003a for details). The results presented thus far were derived with just one iteration. Figure 6 shows the small refinement gained from a second iteration for one of the short exposure datasets (O6IF02010). At low signal levels the simulation derived corrections move closer to the empirical corrections.

Figure 6: As figure 4 but showing the effect deriving the simulation correction from a second iteration.

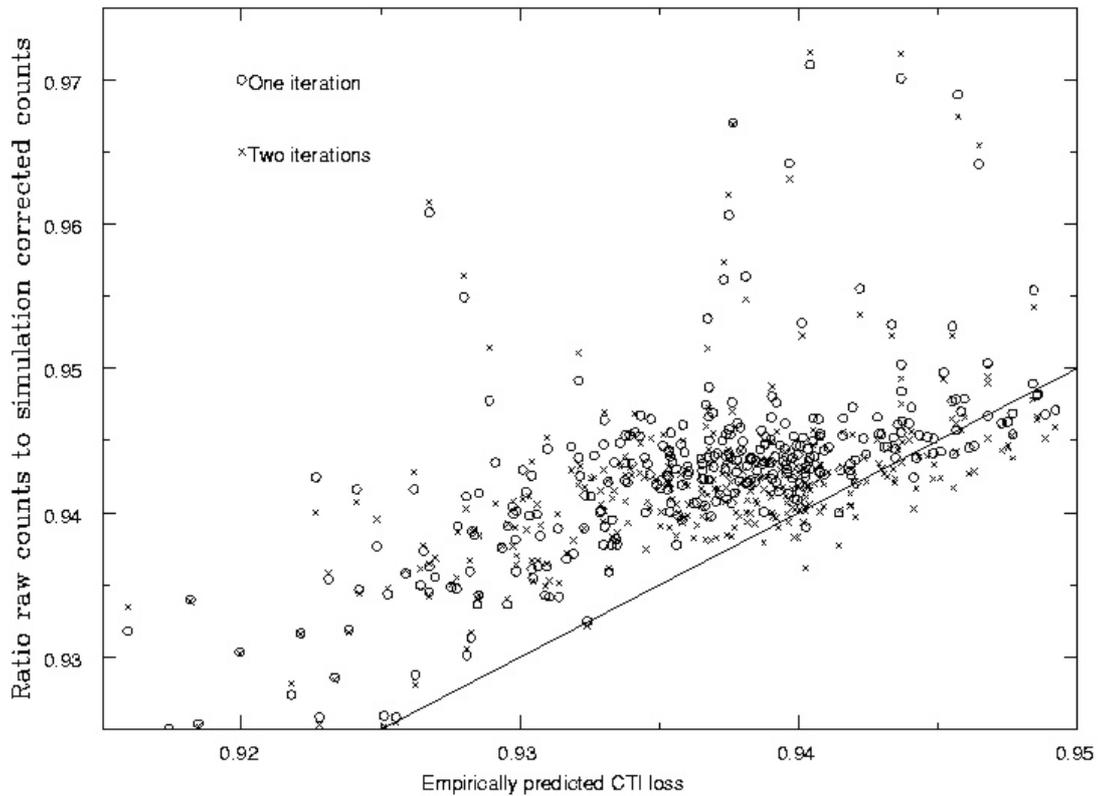

## 5.Summary

Alignment effects and noise make subtle CTI effects in STIS spectroscopic data difficult to demonstrate directly in individual datasets. However, we show the restoration of the vertical cross section of a typical spectrum using simulation derived corrections.

In the process of testing our model, we enhanced the pipeline scripts to perform an automatic application of BG2003's empirical corrections to extracted spectra. We note some ambiguity in the application of these corrections.

Our CCD readout model is able to correct STIS CCD spectroscopic data and produce results consistent with those obtained with the empirically derived corrections of BG2003 for a range of background and signal levels. Moreover, close inspection of discrepancies between simulation derived and empirical corrections reveals advantages in the use of the simulation derived corrections in cases where the charge distribution outside of the extraction area was not simply low level background. This is especially clear where cosmic rays and hot pixels are present but is likely to be responsible for more subtle effects in the case of deviations in the background.